\documentclass[11pt,twoside]{article}
\usepackage{asp2010}

\resetcounters

\begin{document}

\title{Astro-WISE processing of wide-field images and other data}
\author{Hugo Buddelmeijer$^1$, O. Rees Williams$^2$, John P. McFarland$^1$, and Andrey Belikov$^1$
\affil{$^1$Kapteyn Astronomical Institute, Postbus 800, 9700 AV Groningen, The Netherlands}
\affil{$^2$Donald Smits Center for Information Technology, Postbus 11044, 9700 CA Groningen, The Netherlands}
}

\begin{abstract}

{\sf Astro-WISE}\footnote{\url{http://www.astro-wise.org}} \citep{P160_adassxxi} is the Astronomical Wide-field Imaging System for Europe \citep{Astro-WISE}.  
It is a scientific information system which consists of hardware and software federated over about a dozen institutes 
throughout Europe.  It has been developed to exploit the ever increasing avalanche of data produced by astronomical surveys and data intensive scientific experiments in general.

The demo explains the architecture of the {\sf Astro-WISE} information system and shows the use of {\sf Astro-WISE} interfaces.  
Wide-field astronomical images are derived from the raw image to the final catalog according to the user's request.  
The demo  is based on the standard {\sf Astro-WISE} guided tour, which can be accessed from the {\sf Astro-WISE} website.

The typical {\sf Astro-WISE} data processing chain is shown, which can be used for data handling for a variety of different instruments, currently 14, including OmegaCAM, MegaCam, WFI, WFC, ACS/HST, etc\footnote{\url{http://www.astro-wise.org/portal/instruments_index.shtml}}.  

\end{abstract}

\section{Typical Data Processing Chain}

\begin{figure*}[t]
\begin{center}
\includegraphics[angle=0,width=120mm]{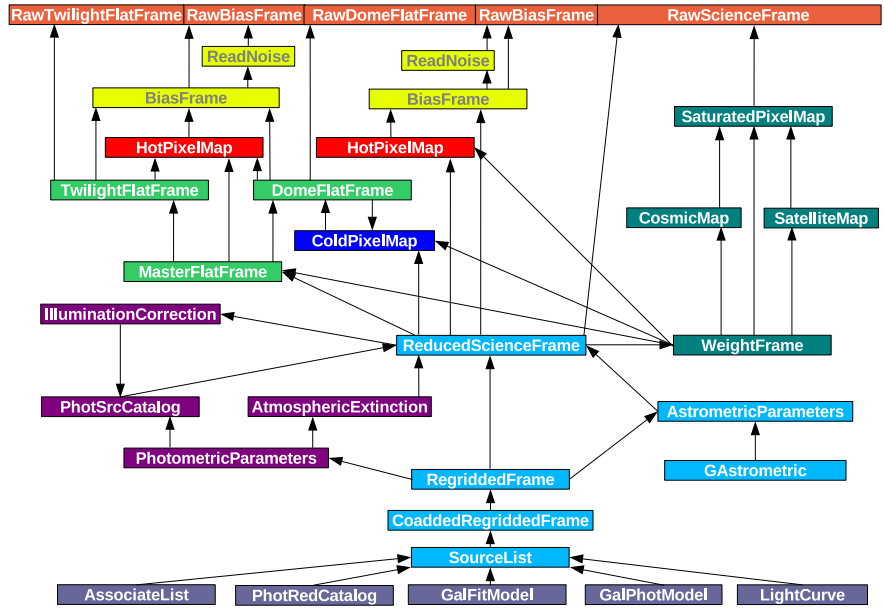}
\caption{
A \textit{target diagram}: slightly simplified object model that is a view of
the dependencies of ``targets'' to the ocean of raw observational data of
astronomical wide-field imaging.  The arrows indicate the backward chaining to
the raw data, not the progression through any processing pipeline.  The colors
provide a visual grouping of similar types of data products.
}\label{fig:target}
\end{center}
\end{figure*}

{\sf Astro-WISE} was originally designed to handle the large datasets from the OmegaCAM instrument such as the KiDS survey \citep{P157_adassxxi}.
The backbone of {\sf Astro-WISE} is set by the way it captures all science products obtained in survey operations in an object-oriented data model.
Figure~\ref{fig:target} shows the main astronomical classes in the {\sf Astro-WISE} environment, the basic elements of the data processing chain.
Each block is a class and each instance of a class is called a \textit{Target}.
The class incorporates a method to derive the data of the Target from other objects, called its \textit{dependencies}.

The user has an ability to combine recipes in a pipeline to process the data by directly requesting the final required data product he or she is interested in.
All processing parameters, along with the full data lineage, are saved in the metadata through the persistence of all objects. 
In the demo the user goes from a {\it RawScienceFrame} (raw frame observed by the VST and ingested in {\sf Astro-WISE}) to a {\it SourceList} (the catalog produced from reduced, regridded and coadded images)\footnote{\url{http://www.astro-wise.org/portal/howtos/man_howto_tutorial_science/man_howto_tutorial_science.shtml}}, using the {\sf Astro-WISE} infrastructure through the web services described below.

An {\sf Astro-WISE} node is the building element of {\sf Astro-WISE} infrastructure.
It consists of data storage element (dataserver, which stores all the files with images), metadata database (RDBMS, which keeps metadata including links between data items), 
computing elements (Distributed Processing Unit) and a number of interfaces and services which allow to the user to browse and process data stored in the system (see~\citet{JOGC} for more technical details).

\section{Services and Interfaces}

The main language for the system is Python, but each user can develop her/his own application or use an existing application 
which can be wrapped into Python. Usually, users develop pipelines or workflows using  existing ``blocks'' with the 
help of pre-defined Python libraries and classes. The user can also change an existing data model if necessary or implement a new one. 

The Command Line Interface of {\sf Astro-WISE}, {\tt AWE} ({\sf Astro-WISE} Environment), can be installed on a site without any other components of {\sf Astro-WISE} (data server and metadata 
database). Basically the {\tt AWE} prompt is a link to a local Python installation with the {\sf Astro-WISE} libraries and environments. 

Apart from the {\tt AWE} prompt, {\sf Astro-WISE} supports a range of web interfaces. This makes it possible for a user to work with data stored in {\sf Astro-WISE} 
without the {\tt AWE} prompt using a web browser only. The web interfaces are divided into two types: data browsing/exploration and data processing/qualification. 
The first group includes:
\begin{itemize}
\item dbviewer\footnote{\url{http://dbview.astro-wise.org}} -- the metadata database interface which allows browsing and querying all attributes of all persistent objects stored in the system,
\item Go-WISE\footnote{\url{http://gowise.astro-wise.org}} -- allows querying on a limited subset of attributes of the data model (coordinate range and object name), and provides results of all projects,
\item image cut out service\footnote{\url{http://cutout.astro-wise.org}} and color image maker\footnote{\url{http://rgb.astro-wise.org}} -- these two services are for 
the astronomical image data type and allows creation of cut-outs of an image or the creation of a pseudo-color RGB image from three different images of the same part of sky,
\item skymap\footnote{\url{http://skymap.astro-wise.org}} -- exploration tool of the {\sf Astro-WISE} system using the GoogleSky interface.
\end{itemize}
Data processing / qualification interfaces are:
\begin{itemize}
\item target processor\footnote{\url{http://process.astro-wise.org}} -- the main web tool to process the data in {\sf Astro-WISE}. This web interface allows the user to go through pre-defined 
processing chains submitting jobs on the {\sf Astro-WISE} computing resources with the ability to select the computing node of {\sf Astro-WISE}.
The Target Processor allows for implicit collaboration by indicating that objects can be reprocessed when another scientist has created improved versions of the objects that it depends on,
\item quality service\footnote{\url{http://quality.astro-wise.org}} --  allows the estimation the quality of the data processing and set a flag highlighting the quality of the data,
\item CalTS\footnote{\url{http://calts.astro-wise.org}} -- web interface for qualifying and time stamping calibration data. 
\end{itemize}

The exact set of web interfaces depends on the data model implemented in the system. The web interfaces described above are for the optical image 
data processing and follow the requirements for this particular data model and data processing chain. {\sf Astro-WISE} allows the implementation 
of new 
web interfaces for the data model and for data processing defined by the user. The developer of the new web interface will use pre-defined 
classes and libraries of {\sf Astro-WISE} to create it.  

\section{Data publishing and External Data}

Data access interfaces from the Virtual Observatory exist as separate services\footnote{\url{http://www.astro-wise.org/portal/aw_vo.shtml}}, that enables browsing the metadata database and retrieving the data from dataservers.
{\sf Astro-WISE} supports the Simple Image Access Protocol for images and ConeSearch for sources.
Each data entity in {\sf Astro-WISE} has a persistent attribute which shows scope of visibility for this entity, which allows the creator of the object to specify with whom to share the object.

\section{Conclusion}
The demo of {\sf Astro-WISE} is based on the {\sf Astro-WISE} guided tour and tutorial, which can be found on the {\sf Astro-WISE} website.
It shows how the request driven way of processing and full data lineage gives {\sf Astro-WISE} the power to handle the large datasets produced by surveys such as KIDS.
The user can apply standard data processing using target processing or can develop his/her own recipe for the data processing using the {\sf Astro-WISE} pipeline as the building blocks. 

\acknowledgements
{\sf Astro-WISE} is an on-going project which started from a FP5 RTD programme funded
by the EC Action ``Enhancing Access to Research Infrastructures''. This work is
supported by FP7 specific programme ``Capacities - Optimising the use and
development of research infrastructures''.
This work is supported by Target\footnote{\url{http://www.rug.nl/target}}, a public-private R\&D programme for information systems for large scale sensor networks. 
\bibliography{D5}

\begin{thebibliography}{}
\expandafter\ifx\csname natexlab\endcsname\relax\def\natexlab#1{#1}\fi
\expandafter\ifx\csname url\endcsname\relax
  \def\url#1{\texttt{#1}}\fi
\expandafter\ifx\csname urlprefix\endcsname\relax\def\urlprefix{URL }\fi
\providecommand{\eprint}[2][]{\url{#2}}

\bibitem[{Begeman et~al.(2010)Begeman, Belikov, Boxhoorn, Dijkstra, Valentijn,
  Vriend, \& Zhao}]{JOGC}
Begeman, K., Belikov, A., Boxhoorn, D., Dijkstra, F., Valentijn, E., Vriend,
  W.-J., \& Zhao, Z. 2010, Journal of Grid Computing, 8, 199

\bibitem[{Valentijn et~al.(2007)Valentijn, McFarland, Snigula, Begeman,
  Boxhoorn, Rengelink, Helmich, Heraudeau, Verdoes~Kleijn, Vermeij, Vriend,
  Tempelaar, Deul, Kuijken, Capacciolo, Silvotti, Bender, Neeser, Saglia,
  Bertin, \& Mellier}]{Astro-WISE}
Valentijn, E., McFarland, J., Snigula, J., Begeman, K., Boxhoorn, D.,
  Rengelink, R., Helmich, E., Heraudeau, P., Verdoes~Kleijn, G., Vermeij, R.,
  Vriend, W.-J., Tempelaar, M., Deul, E., Kuijken, K., Capacciolo, M.,
  Silvotti, R., Bender, R., Neeser, M., Saglia, R., Bertin, E., \& Mellier, Y.
  2007, in Proc. of ADASS XVI, edited by B.~D. Shaw~R.A., Hill~F., vol. 376 of
  ASP Conf. Ser., 491

\bibitem[{Verdoes~Kleijn et~al.(2012)Verdoes~Kleijn, de~Jong, Valentijn,
  Kuijken, Bout, Boxhoorn, Helmich, McFarland, \& Sikkema}]{P157_adassxxi}
Verdoes~Kleijn, G.~A., de~Jong, J., Valentijn, E.~A., Kuijken, K., Bout, J.,
  Boxhoorn, D., Helmich, E., McFarland, J., \& Sikkema, G. 2012, in ADASS XXI,
  edited by P.~Ballester, \& D.~Egret (San Francisco: ASP), vol. TBD of ASP
  Conf. Ser., TBD

\bibitem[{Vriend et~al.(2012)Vriend, Valentijn, Belikov, \&
  Verdoes~Kleijn}]{P160_adassxxi}
Vriend, W.-J., Valentijn, E.~A., Belikov, A.~N., \& Verdoes~Kleijn, G.~A. 2012,
  in ADASS XXI, edited by P.~Ballester, \& D.~Egret (San Francisco: ASP), vol.
  TBD of ASP Conf. Ser., TBD

\end{thebibliography}

\end{document}